\documentclass[twocolumn,showpacs,preprintnumbers,amsmath,amssymb]{revtex4-1}
\usepackage{graphicx}
\usepackage{bm}
\usepackage{amsmath}
\usepackage{multirow}
\usepackage{txfonts}
\begin{document}

\title{Time-reversal-breaking topological phases in antiferromagnetic Sr$_2$FeOsO$_6$ films}
\author{Xiao-Yu Dong$^{1,2}$, Sudipta Kanungo$^{3,4}$, Binghai Yan$^{2,3}$, Chao-Xing Liu$^{5}$}
\affiliation{$^{1}$Department of Physics and State Key Laboratory of Low-Dimensional Quantum Physics, Tsinghua University, Beijing 100084, P.R.China;}
\affiliation{$^{2}$Max-Planck-Institut f\"ur Physik komplexer Systeme, 01187, Dresden, Germany;}
\affiliation{$^{3}$Max-Planck-Institut f\"ur Chemische Physik fester Stoffe, 01187 Dresden, Germany;}
\affiliation{$^{4}$Center for Emergent Matter Science (CEMS), RIKEN, 2-1, Hirosawa, Wako, Saitama 351-0198, Japan}
\affiliation{$^{5}$Department of Physics, The Pennsylvania State University, University Park, Pennsylvania 16802-6300, USA;}
\date{\today}

\begin{abstract}
In this work, we studied time-reversal-breaking topological phases as a result of the interplay between antiferromagnetism and inverted band structures in antiferromagnetic double perovskite transition-metal Sr$_2$FeOsO$_6$ films. By combining the first-principles calculations and analytical models, we demonstrate that the quantum anomalous Hall phase and chiral topological superconducting phase can be realized in this system. We find that to achieve time-reversal-breaking topological phases in antiferromagnetic materials, it is essential to break the combined symmetry of time reversal and inversion, which generally exists in antiferromagnetic structures. As a result, we can utilize an external electric gate voltage to induce the phase transition between topological phases and trivial phases, thus providing an electrically controllable topological platform for the future transport experiments.
\end{abstract}

\maketitle

\section{Introduction}
%{\it Introduction -}
Recent years have witness the rapid development of the field of topological states of matter, due to its importance in our understanding on quantum states of matter, as well as the potential applications in electronic devices with low dissipations \cite{RevModPhys.83.1057,RevModPhys.82.3045} and topological quantum computation \cite{kitaev2003fault,RevModPhys.80.1083}. The classification of topological states depends on the presence or absence of certain type of symmetry \cite{schnyder2008classification,ryu2010topological}. Quantum anomalous Hall (QAH) phase \cite{qi2006topological,yu2010quantized} and chiral topological superconductor (CTSc) \cite{alicea2012new} are two examples of topological states in absence of time reversal (TR) symmetry (denoted as $\hat{T}$). The realization of these time-reversal-breaking (TRB) topological states in ferromagnetic (FM) materials, as a consequence of the interplay of exchange coupling, spin-orbit coupling (SOC) and inverted band structures, has motivated intensive theoretical and experimental research activities. The QAH effect has been observed experimentally in magnetically doped (Bi,Sb)$_2$Te$_3$ films with long-range ferromagnetic order \cite{chang2013experimental,chang2015high}. Evidences of CTSc have been identified in Sr$_2$RuO$_4$ $p$-wave superconductors (SCs) \cite{RevModPhys.75.657} or other SC system in proximity to FM or under magnetic fields \cite{mourik2012signatures,nadj2014observation}.
%semiconducting nano-wires in the proximity to superconductors under an external magnetic field \cite{mourik2012signatures}, as well as ferromagnetic atom chains on top of superconductor substrates \cite{nadj2014observation}.
Since any magnetism can break TR symmetry, it is natural to ask if TRB topological phases can be realized in other magnetic structures, in particular anti-ferromagnetic (AFM) materials. A few recent theoretical works started exploring topological phases in AFM systems \cite{1367-2630-15-6-063031,wu2015quantum,zhou2016two,qiao2014quantum,PhysRevLett.116.256601}. However, the required condition is still unclear. More importantly, superconductivity is incompatible with ferromagnetism, but it can coexist with anti-ferromagnetism \cite{bennemann2008superconductivity}. Thus, our understanding of the interplay between topologically non-trivial band structures and anti-ferromagnetism is important for the search of new topological superconducting materials.

In this work, we studied TRB phases in a two dimensional (2D) AFM transition metal perovskite material -- Sr$_2$FeOsO$_6$ thin film -- on top of insulating or superconducting substrates. With a realistic tight-binding model, we demonstrate that both the anomalous Hall (AH) and CTSc states can be realized in this material by applying an external electric gate voltage to the film. The combined  $\hat{T}\hat{P}$ symmetry, where $\hat{P}$ denotes inversion symmetry, plays an essential role since it can protect a double degeneracy at each momentum in the Brillouin zone (BZ), and forbid the presence of the AH and CTSc states. Since the $\hat{T}\hat{P}$ symmetry can be broken by an external electric gate voltage, this allows for an electric switch between different TRB topological states. We map out the phase diagram of various CTSc phases as a function of external gate voltages and chemical potential, based on which one can construct junction structures to detect chiral Majorana fermions.

% By tuning the external electronic gate voltage, a band inversion happens, and this lead to a QAH phase. The superconducting pairing terms are introduced by the proximity effect when coupled to an $s$-wave superconductor. TSc phases could be generated from the QAH state, as proposed by Xiao-Liang Qi et.al \cite{PhysRevB.82.184516}. Moreover, with a gate voltage that is much smaller than the gate voltage used above to induce band inversion, a TSc phase could appear due the existence of a single hole pocket and spin-orbit coupling. Tuning the gate voltage and chemical potential, various TSc phases could be obtained.

\section{S\lowercase{r}$_2$F\lowercase{e}O\lowercase{s}O$_6$ films}
\begin{figure} %[h]
\includegraphics[width=8.6cm]{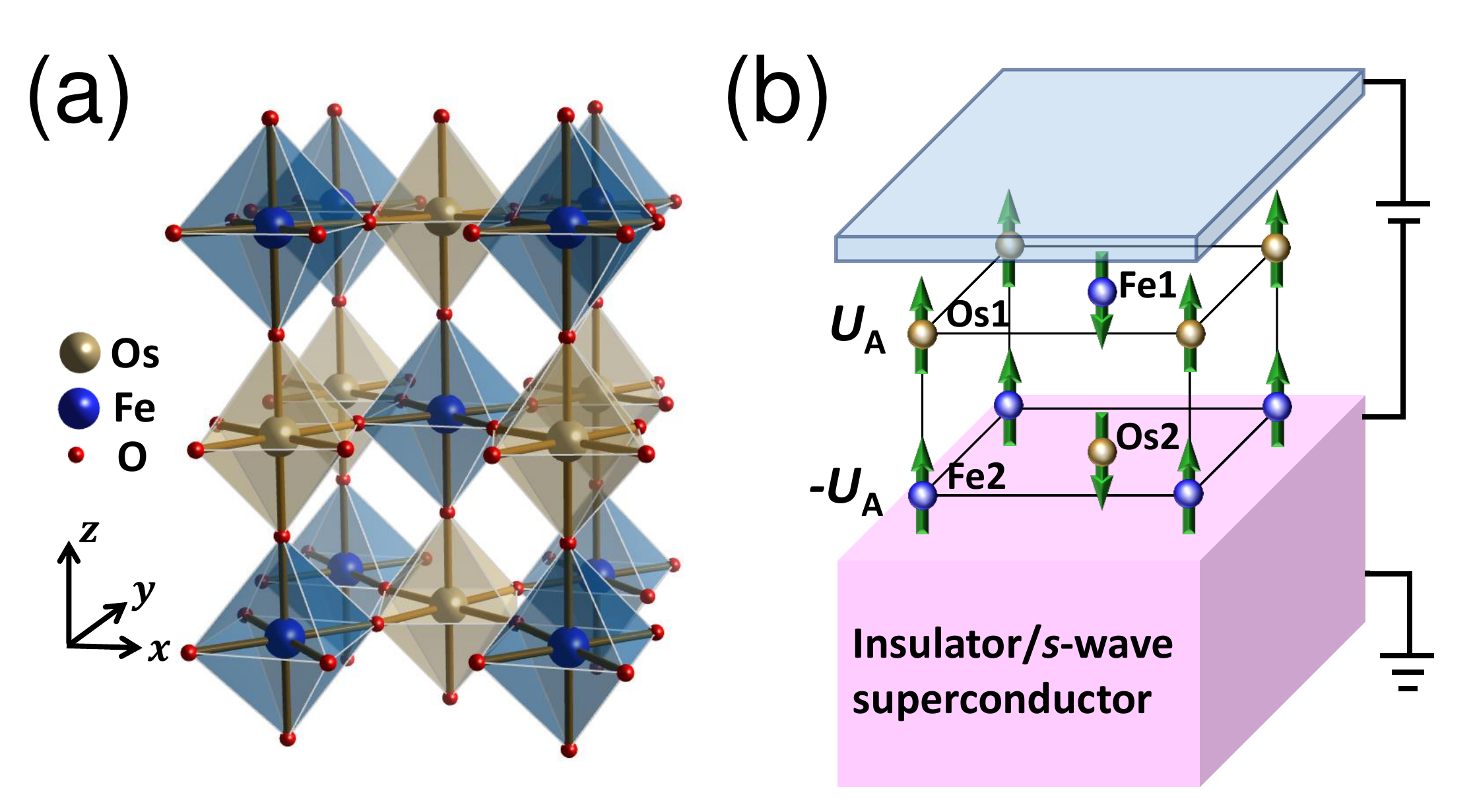}
\caption{(a) The crystal structure of Sr$_2$FeOsO$_6$. One tetragonal unit cell is shown, where the Sr atoms are omitted for simple.
 (b) Schematic plot of the bilayer film with gate voltage $U_{\mathrm{A}}$. The arrows on the atoms denote the local magnetic moments. The substrate could be normal insulator or $s$-wave superconductor.}   \label{Fig:1}
\end{figure}

\begin{figure} %[h]
\includegraphics[width=8.6cm]{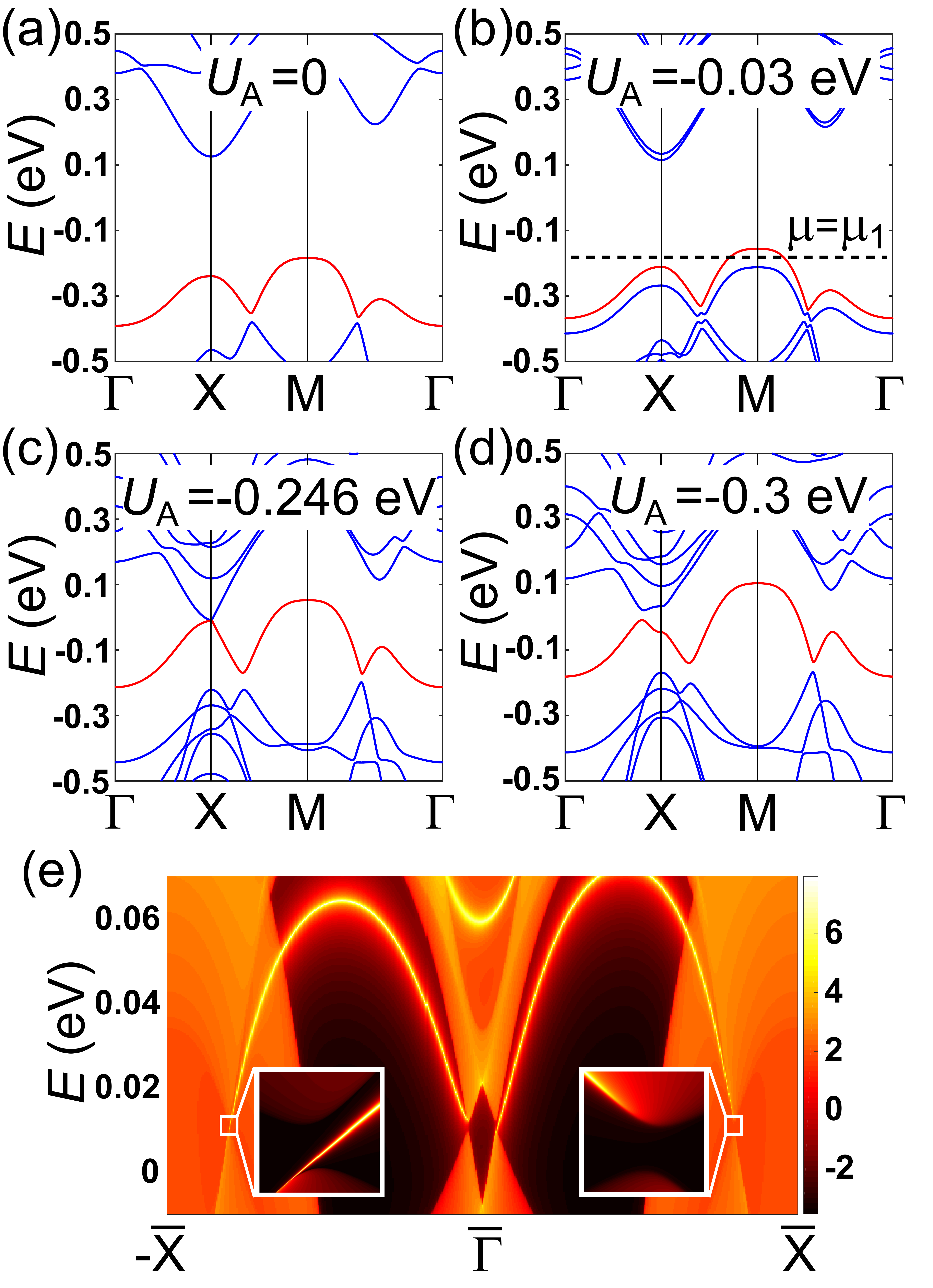}
\caption{(a)--(d) The band structure of the bilayer film with $U_{\mathrm{A}}=0,-0.03,-0.246,-0.3$ eV, respectively. (e) The density of states on one edge of a ribbon configuration along $x$ direction with $U_{\mathrm{A}}=-0.3$ eV. Two insets are the zoom-in of the corresponding regions.}   \label{Fig:2}
\end{figure}

%{\it Sr$_2$FeOsO$_6$ films -}
The material Sr$_2$FeOsO$_6$ belongs to a family of transition metal double perovskites $A_2BB'\mathrm{O}_6$, in which $A$ could be an alkali, alkaline earth, or rare earth atom, while $B$ and $B'$ are two different transition metal atoms \cite{PhysRevLett.111.167205}. As shown in Fig. \ref{Fig:1}(a), % ({\bf BH, please add a figure for the crystal.}).
Sr$_2$FeOsO$_6$ crystalizes into an well-order tetragonal lattice with the $\mathrm{FeO}_6$ and $\mathrm{OsO}_6$ forming corner-sharing octahedra.
%({\bf BH, please check if the statement is accurate or not and if we need to talk anything more about AFM phase for this system})
Recent experiments reveals \cite{paul2013synthesis,PhysRevLett.111.167205,feng2013high} that Sr$_2$FeOsO$_6$ is an AFM semiconductor with two magnetic phase transitions at 140 K and 67 K. Both the higher and lower temperature phases are AFM, in which magnetic moments of Fe ($S=5/2$) and Os ($S=3/2$) atoms align along the $z$ direction according to the neutron diffraction~\cite{PhysRevLett.111.167205}.
Magnetic sites Fe and Os couple in an AFM way inside one checkerboard atomic layer in the $xy$ plane and two adjacent atomic layers form a double layer configuration by the FM coupling.
Further FM coupling between neighboring double layers along the $z$ axis leads to the higher temperature phase, while the AFM coupling between double layers gives rise to the lower temperature phase \cite{PhysRevLett.111.167205,kanungo2014ab}.
%{\color{blue}[also Phys. Rev. B 89, 214414 (2014)]} ({\bf XY, add this reference}).
In this work, we consider the principle double layers of Sr$_2$FeOsO$_6$ with out-of-plane magnetic moments aligned in the in-plane AFM ordering, as shown in Fig. \ref{Fig:1}(b). Low energy physics are dominated by the $d$ orbitals of Fe and Os atoms, and thus we only focus on these two atoms, which form a layered checkerboard lattice for each layer.
 %formed by Fe and Os atoms. Fe atom in one layer holds the same position in $xy$-plane with Os atom in the adjacent layer. The anti-ferromagnetic ordering in the bilayer structure is shown in
For a bilayer lattice, there are four atoms in one unit cell, denoted as $\mathrm{Fe}_1$ and $\mathrm{Os}_1$ in the top layer and $\mathrm{Fe}_2$ and $\mathrm{Os}_2$ in the bottom layer. The magnetic moments of $\mathrm{Os}_1$ and $\mathrm{Fe}_2$ atoms
%, which occupy the same position of the checkerboard lattice in the $xy$ plane,
point up along the $z$ direction, while those of $\mathrm{Fe}_1$ and $\mathrm{Os}_2$ atoms point down (Fig. \ref{Fig:1}(b)).
Based on the maximum localized Wannier function method \cite{marzari1997maximally,souza2001maximally}, we construct a realistic tight-binding Hamiltonian, labeled by $H_0({\bf k})$, with all five $d$-orbitals of Fe atoms and the $t_{2g}$ part (three orbitals) of Os $d$-orbitals. We justify our tight-binding model by comparing the bulk energy dispersion with that from the first principles calculations and a good agreement is found, as shown in the Appendix \ref{App:A}. The eigen-energy spectrum  of the bilayer system can be obtained by solving the eigen equation $H_0({\bf k})|\psi_n\rangle=E_n|\psi_n\rangle$ of this $32\times 32$ Hamiltonian $H_0({\bf k})$ (16 orbitals and two spins), as shown in Fig. \ref{Fig:2}(a), for external gate voltage $U_{\mathrm{A}}=0$. Here the potential induced by an external gate voltage is assumed to be $U_{\mathrm{A}}$ on the top layer and $-U_{\mathrm{A}}$ on the bottom layer.
%The electronic band structure of the bilayer film is depicted in Fig. \ref{Fig:2}(a) based on the calculation of a tight-binding model built with the maximum localized Wannier function method (five $d$-orbits of Fe and $d_{xz}$, $d_{yz}$, and $d_{xy}$ orbits of Os have been used to build the model).
From Fig. \ref{Fig:2}(a), we find that the conduction band minimum appears at X$=(\pi,0)$ (or equivalent Y$=(0,\pi)$) while the valence band maximum at M$=(\pi,\pi)$ is about tens of meV higher than that at X (or Y), giving rise to an indirect band gap. An intriguing feature is that all the bands are doubly degenerate at each momentum, similar to the case of Kramer's degeneracy in a TR invariant system \cite{dresselhaus2007group}. Although there is no individual $\hat{T}$ or $\hat{P}$ symmetry, the combined symmetry $\hat{T}\hat{P}$ exists, which reverses spin and interchanges layers, but preserves momentum, thus leading to the double (spin) degeneracy. This is quite similar to the Kramer's degeneracy, and forbids the occurrence of non-zero Hall conductance (or equivalently non-zero Chern number). Therefore, it is essential to split this double degeneracy by breaking $\hat{T}\hat{P}$ symmetry to achieve any TRB topological phase.
%We find the local maxima of the valence bands at both M$=(\pi,\pi)$ and X$=(0,\pi)$ (or equivalent Y$=(\pi,0)$ due to four-fold rotation symmetry around $z$ axis) and the valence band maximum at M is slightly higher than that at X (Y).
We notice this degeneracy can be split by introducing an electric field along the $z$ direction to break $\hat{T}\hat{P}$ symmetry.
%that two degenerate states come from opposite layers because $\hat{T}\hat{P}$ symmetry operation interchanges two layers. We may introduce opposite potentials ($U_{\mathrm{A}}$) into two layers, which can be easily induced by a $z$-directional electric field in experiments, to split this degeneracy by breaking $\hat{T}\hat{P}$ symmetry.
Below we will demonstrate that this strategy allows us to realize the AH phase and CTSc phase in thin films of Sr$_2$FeOsO$_6$.

%$\hat{I}\hat{T}$ symmetry could be broken explicitly by applying an asymmetry electric potential through an external gate on the film. As a result of symmetry breaking, the double degeneracy will be lost, as shown in Fig. 2(b). Assuming the electric gate potential $V_g>0$, that is the effective electric potential in the upper layer is higher than that in the lower layer, the energy of the state $|\psi_{1}\rangle_\mathrm{M}$ will be pushed up and the energy of $|\psi_{2}\rangle_\mathrm{M}$ will be pushed down, relatively. In this way, we get a single hole pocket around M, which is very important for the realization of a TSc phase that will discussed later in this paper.

\section{Anomalous Hall phase}
%{\it Quantum Anomalous Hall phase} ---
We first study the case without superconductivity. % and study the influence of a strong asymmetric potential along the $z$-direction induced by a gate voltage.
%Since the direct band gap exists only at X (Y) (Fig. \ref{Fig:2}(a)), %(the conduction band energy at M is high),
We only focus on the X and Y points with a direct band gap (Fig. \ref{Fig:2}(a)). By applying a non-zero asymmetric potential $U_{\mathrm{A}}$ (Fig. \ref{Fig:2}(b), (c) and (d)), double degeneracy for both conduction and valence bands are split. This splitting greatly reduces band gap (Fig. \ref{Fig:2}(b) and (c)), and even reverses band ordering at X (Fig. \ref{Fig:2}(d)), leading to band inversion. Band inversion can result in topological phase transition in the field of topological insulators \cite{RevModPhys.83.1057,RevModPhys.82.3045}. Thus, we expect that band structure in Fig. \ref{Fig:2}(d) is topologically nontrivial. We can use Chern number, defined as $C=(1/2\pi)\int dk_x\int dk_y (\partial_{k_x}A_y(\mathbf{k})-\partial_{k_y}A_x(\mathbf{k}))$, where $A_i=-i\sum_{n\in \mathrm{occ.}}\langle \psi_n(\mathbf{k})|\partial_{k_i}|\psi_n(\mathbf{k}) \rangle$ (the summation is taken over all of the occupied bands), to characterize topological nature of our system, and direct calculation shows that the Chern number $C$ is $0$ for $U_{\mathrm{A}}>-0.246$ eV and $+2$ for $U_{\mathrm{A}}<-0.246$ eV. Furthermore, we study the low-energy effective theory.
Let us label two valence bands at X as $|\psi_{\mathrm{v}1}\rangle_{\mathrm{X}}$ and $|\psi_{\mathrm{v}2}\rangle_{\mathrm{X}}$ and two conduction bands as $|\psi_{\mathrm{c}1}\rangle_{\mathrm{X}}$ and $|\psi_{\mathrm{c}2}\rangle_{\mathrm{X}}$.
The first principles calculations show that $|\psi_{\mathrm{v}1}\rangle_{\mathrm{X}}$ originates from $|\mathrm{Os}_1,d_{yz},\uparrow\rangle$, $|\psi_{\mathrm{v}2}\rangle_{\mathrm{X}}$ is dominated by $|\mathrm{Os}_2,d_{yz},\downarrow\rangle$, $|\psi_{\mathrm{c}1}\rangle_{\mathrm{X}}$ mainly consists of $|\mathrm{Os}_1,d_{xz},\downarrow\rangle$, and $|\psi_{\mathrm{c}2}\rangle_{\mathrm{X}}$ is characterized by $|\mathrm{Os}_2,d_{xz},\uparrow\rangle$.
%At M, $|\psi_{\mathrm{v}1}\rangle_{\mathrm{M}}$ is dominated by the spin-up $d_{xy}$-orbital state of $\mathrm{Os}_1$ atoms, denoted as $|\mathrm{Os}_1,d_{xy},\uparrow\rangle$, while $|\psi_{\mathrm{v}2}\rangle_{\mathrm{M}}$ is dominated by $|\mathrm{Os}_2,d_{xy},\downarrow\rangle$ according to the $\hat{T}\hat{I}$ symmetry.
%Let us label the two degenerate highest valence band states at M (X) as $|\psi_{\mathrm{v}1}\rangle_{\mathrm{M(X)}}$ and $|\psi_{\mathrm{v}2}\rangle_{\mathrm{M(X)}}$. At M, $|\psi_{\mathrm{v}1}\rangle_{\mathrm{M}}$ is dominated by the spin-up $d_{xy}$-orbital state of $\mathrm{Os}_1$ atoms, denoted as $|\mathrm{Os}_1,d_{xy},\uparrow\rangle$, while $|\psi_{\mathrm{v}2}\rangle_{\mathrm{M}}$ is dominated by $|\mathrm{Os}_2,d_{xy},\downarrow\rangle$ according to the $\hat{T}\hat{I}$ symmetry. At X, $|\psi_{\mathrm{v}1}\rangle_{\mathrm{X}}$ is dominated by $|\mathrm{Os}_1,d_{yz},\uparrow\rangle$, while $|\psi_{\mathrm{v}2}\rangle_{\mathrm{X}}$ is dominated by $|\mathrm{Os}_2,d_{yz},\downarrow\rangle$.
%The magnetic moments on $\mathrm{Os}_1$ and $\mathrm{Os}_2$ are opposite with the same absolute value, thus $|\mathrm{Os}_1,d_{xy},\uparrow\rangle$ and $|\mathrm{Os}_2,d_{xy},\downarrow\rangle$ should have the same energy.
When the asymmetric potential $U_{\mathrm{A}}$ is applied, the band inversion occurs between $|\psi_{\mathrm{c}1}\rangle_{\mathrm{X}}$ and $|\psi_{\mathrm{v}2}\rangle_{\mathrm{X}}$. On the basis of $\Psi_{\mathrm{X}}=(|\psi_{\mathrm{c}1}\rangle_\mathrm{X}, |\psi_{\mathrm{v}2}\rangle_\mathrm{X})^T$, the effective Hamiltonian is
%Since the low energy physics is dominated by these two bands, this allows us to write down a two-band effective model, described by the Hamiltonian
%Due to the double perovskites lattice structure, in which two ferromagnetic sublattices in the $xy$-plane with opposite magnetic moments are constructed by different atoms, the in-plane mirror symmetries are broken, thus the QAH phases are expected to exist. Now, we neglect the hole pocket around M for a moment, and concentrate on the properties around X. Increasing the gate voltage, the band gap at X will close and reopen again (a band inversion happens around X), as shown in Fig. 2(c) and (d). Around the gapless region the bands disperse linearly. With a positive gate voltage $V_{\mathrm{g}}>0$, the lowest conduction band at X ($|\psi_1\rangle_\mathrm{X}$) is dominated by $d_{xz}$ orbit of Os$_2$ in the lower layer with spin up, denoted as $|\mathrm{Os}_2,d_{xz},\uparrow\rangle$; while the highest valence band at X ($|\psi_2\rangle_\mathrm{X}$) is dominated by $d_{yz}$ orbit of Os$_1$ in the upper layer with spin up, denoted as $|\mathrm{Os}_1,d_{yz},\uparrow\rangle$. In the basis $\Psi_{\mathrm{X}}=(|\psi_1\rangle_\mathrm{X}, |\psi_2\rangle_\mathrm{X})^T$, the two-band effective model around X is
\begin{eqnarray}
  H_{\mathrm{eff}}&=&\epsilon(\mathbf{k})\sigma^0+m(\mathbf{k})\sigma^z +l_1 k_x\sigma^x+l_2 k_y\sigma^y,
\end{eqnarray}
where $\epsilon({\bf k})=a_0+a_1k_x^2+a_2k_y^2$, $m({\bf k})=m_0+m_1k_x^2+m_2k_y^2$ and $a_{0,1,2}, m_{0,1,2}, l_{1,2}$ are material dependent parameters. We recognize this model as a 2D massive Dirac Hamiltonian and the band inversion, which is described by the sign reverse of the mass $m_0$, changes the Chern number $C$ by $\pm 1$. Therefore, the Chern number $C$ for $U_{\mathrm{A}}<-0.246$ eV should be $+2$, taking into account the band inversion at both X and Y. Thus, we conclude that the Hall conductance should be $2e^2/h$ for $U_{\mathrm{A}}<-0.246$ eV. This conclusion is further supported by the direct calculation of edge states in the ribbon configuration based on the iterative Green function method \cite{sancho1985highly}. Indeed, we find two chiral edge states at one edge of the ribbon (See Fig. \ref{Fig:2}(e) and its inset).
%One can see that one edge mode appears at the valence band top near $-\bar{\mathrm{X}}$ (see the left inset in Fig. \ref{Fig:2}(e)) and merges to the conduction band bottom around $\bar{\Gamma}$, while another edge mode connects the valence band top near $\bar{\Gamma}$ to the conduction band bottom close to $\bar{\mathrm{X}}$ (see the right inset in Fig. \ref{Fig:2}(e)).
Since the valence band maximum at M is higher than that at X or Y,
electrons may transfer from valence bands at M to conduction bands at X and Y, leading to electron pockets at X and Y and hole pockets at M. The present of bulk carriers will destroy the quantization of Hall resistance.
Thus, a large, instead of quantized, Hall resistance is expected in experiments. In addition, the scattering between electron pockets at X and Y may lead to charge (or spin) density wave. But we emphasize that topological nature of the system will not be changed once the band gap remains under charge density wave (see Appendix \ref{App:B}). Although we consider bilayer film for simplicity, our results also exist for other film thickness (see the calculation for four-layer films in the Appendix \ref{App:C}). We emphasize that splitting spin degeneracy by breaking the $\hat{T}\hat{P}$ symmetry is essential and this strategy can be applied to other anti-ferromagnetic system.

 %Unfortunately, due to the hole pocket around M, the bulk of the 2D film in the band inverted region is gapless, although there is a direct gap at X. Therefore, it is not a good QAH system. We have shown that in principle the QAH phases could exist in an anti-ferromagnetic system with double perovskite lattice structure. We could expect that a good QAH phase with insulating bulk could be found in this family of materials.

\section{Chiral Topological Superconducting phase}
%{\it Chiral Topological Superconducting phase}---
Next we will study this system on top of a superconducting substrate with the $s$-wave singlet pairing and search for CTSc. The key idea is to induce effective triplet SC from singlet SC as a result of the coexistence of magnetism and SOC in the Sr$_2$FeOsO$_6$ film. Similar idea has been applied to CTSc in half-metals in proximity to a SC \cite{lee2009proposal,PhysRevB.84.060510}, where interfacial Rashba SOC flips electron spin and converts singlet pairing to triplet pairing. The advantage here is that anti-ferromagnetism and SOC coexist for Os atoms and can naturally lead to pairing conversion. Below we will demonstrate that CTSc can be realized and controlled by tuning asymmetric potential $U_{\mathrm{A}}$ and chemical potential $\mu$.

To simulate superconducting proximity effect, we consider the Bogoliubov-de Gennes (BdG) Hamiltonian
\begin{eqnarray}
  H_{\mathrm{BdG}}=\left(
  \begin{array}{cc}
    H_0(\mathbf{k})-\mu & i \sigma^y\otimes\tilde{\Delta}\\
    -i\sigma^y\otimes \tilde{\Delta} & -H^T_0(-\mathbf{k})+\mu\\
  \end{array}
  \right)\label{Eq:1}
\end{eqnarray}
in which $H_0$ is the tight-binding model, $\tilde{\Delta}$ is a diagonal matrix with non-zero diagonal elements $\Delta$ to describe the proximity induced $s$-wave pairing for $d$-orbitals of Fe$_2$ and Os$_2$ atoms at the bottom layer. We also define a Chern number, $N=(1/2\pi)\int dk_x\int dk_y (\partial_{k_x}\tilde{A}_y(\mathbf{k})-\partial_{k_y}\tilde{A}_x(\mathbf{k}))$, where $\tilde{A}_i=-i\sum_{n\in \mathrm{hole}}\langle \varphi_n(\mathbf{k})|\partial_{k_i}|\varphi_n(\mathbf{k}) \rangle$, to describe topological property of the BdG Hamiltonian, where $|\varphi_n\rangle$ is eigen wave functions of $H_{\mathrm{BdG}}$ and the summation is taken over all of the hole bands. The Chern number $N$ for $H_{\mathrm{BdG}}$ is not the same as $C$ for $H_0$. For the pairing potential $\Delta=0$, one can show that the electron part $H_0({\bf k})$ and the hole part $-H^T_0(-{\bf k})$ possess the same Chern number $C$ and thus $N=2C$ \cite{PhysRevB.82.184516}. When $\Delta\neq 0$, there is no simple relation between $C$ and $N$.

%Below, we will first present our numerical simulations on the BdG Hamiltonian for bilayer Sr$_2$FeOsO$_6$ film with the on-site $s$-wave superconducting pairing term $\Delta$ for all the orbitals of the atoms Fe$_2$ and Os$_2$ in the bottom layer and discuss the phase diagram of superconducting phases in our system as a function of asymmetric potential $U_{\mathrm{A}}$ and chemical potential $\mu$. To understand the phase diagram from our numerical simulation, we will present our low energy effective theory and argue that it is again essential to break the $\hat{T}\hat{I}$ symmetry, which allows us to tune CTSc phase in an electric manner. Finally, we will discuss possible transport experiments in a junction structure with four-terminal measurements.

\begin{figure} %[h]
\includegraphics[width=8.6cm]{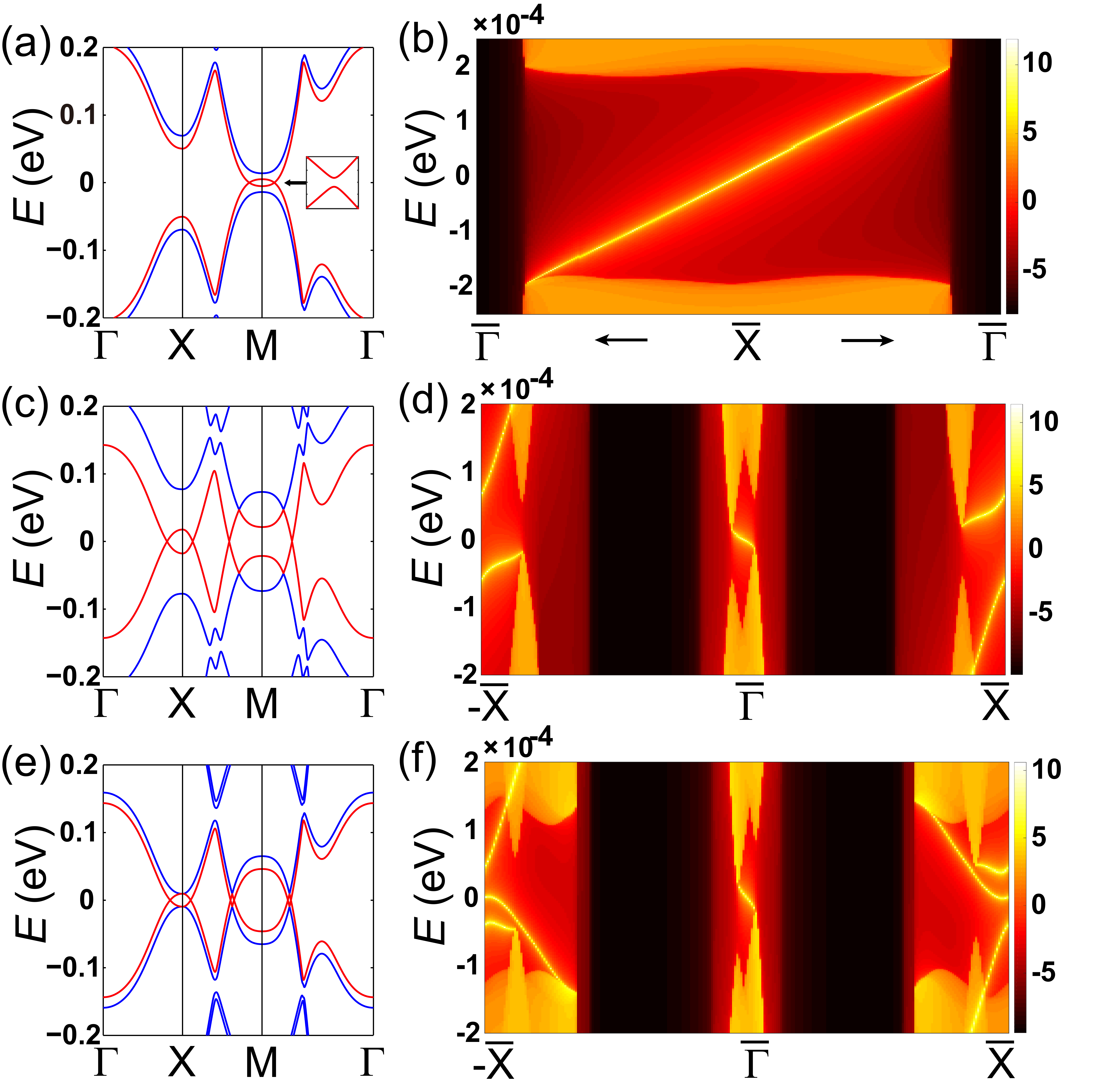}
\caption{(a) The dispersion of the BdG Hamiltonian of the bilayer system with $U_{\mathrm{A}}=-0.01$ eV and $\mu=-0.18$ eV. (b) The edge dispersion in a 1D ribbon configuration along $x$ direction with the same parameters in (a). Correspondingly, the parameters in (c) and (d) are $U_{\mathrm{A}}=-0.05$ eV and $\mu=-0.21$ eV, and in (e) and (f) are $U_{\mathrm{A}}=-0.01$ eV and $\mu=-0.24$ eV. The value of $\Delta$ is fixed at $0.015$ eV in all these figures.}   \label{Fig:3}
\end{figure}

Eigen energies of $H_{BdG}$ can be extracted by solving the eigen-equation (\ref{Eq:1}) and a typical energy dispersion is shown in Fig. \ref{Fig:3}(a). The Fermi energy is tuned to the valence bands due to the rich physics in this regime. Due to the particle-hole symmetry of $H_{\mathrm{BdG}}$, the electron bands are symmetric to the corresponding hole bands with respect to Fermi energy (corresponding to zero energy in Fig. \ref{Fig:3}(a) for $H_{\mathrm{BdG}}$). A superconducting gap induced by $\Delta$ is shown in the inset of Fig. \ref{Fig:3}(a). The key to realize CTSc is to identify the parameter regime in which band crossings between electron and hole bands only occur for odd number of times. When $U_{\mathrm{A}}=0$, all the bands are doubly degenerate due to the $\hat{T}\hat{P}$ symmetry and the number of band crossing must be even. Therefore, the resulting superconducting phase must be trivial. This argument again suggests that breaking $\hat{T}\hat{P}$ symmetry by a non-zero $U_{\mathrm{A}}$ is essential. Fig. \ref{Fig:3}(a), (c) and (e) are for three different sets of parameters ($U_{\mathrm{A}}$ and $\mu$). In Fig. \ref{Fig:3}(a) with $U_{\mathrm{A}}=-0.01$ eV, $\mu=-0.18$ eV and $\Delta=0.015$ eV, a non-zero $U_{\mathrm{A}}$ split two bands at M and the Fermi energy is tuned to cross only one band around M. In this case, the Chern number $N=+1$ for the corresponding superconducting phase with one chiral Majorana edge mode, as confirmed by the direct calculation of edge modes on a slab configuration in Fig. \ref{Fig:3}(b).
%, one chiral Majorana edge mode is clearly found at $\bar{\mathrm{X}}$ ($\bar{\mathrm{X}}$ corresponds to the projection of M to the boundary BZ).
We can further construct an effective model for this superconducting phase based on the theory of invariant. We label two valence band states at M as $|\psi_{\mathrm{v}1}\rangle_{\mathrm{M}}$ and $|\psi_{\mathrm{v}2}\rangle_{\mathrm{M}}$, of which $|\psi_{\mathrm{v}1}\rangle_{\mathrm{M}}$ is dominated by the $|\mathrm{Os}_1,d_{xy},\uparrow\rangle$ state, while $|\psi_{\mathrm{v}2}\rangle_{\mathrm{M}}$ is given by $|\mathrm{Os}_2,d_{xy},\downarrow\rangle$. For $U_{\mathrm{A}}=0$, only one term $(c_0+c_1(k_x^2+k_y^2))\sigma^0$ is allowed (See Appendix \ref{App:D}), where $c_0$ and $c_1$ are material dependent parameters. For a non-zero $U_{\mathrm{A}}$, the two-band model is
\begin{eqnarray}
  h(\mathbf{k})&=&\epsilon(\mathbf{k})\sigma^0 +\epsilon'(\mathbf{k})\sigma^z+t_1(k_x\sigma^x+k_y\sigma^y)
\end{eqnarray}
on the basis of $|\psi_{\mathrm{v}1}\rangle_{\mathrm{M}}$ and $|\psi_{\mathrm{v}2}\rangle_{\mathrm{M}}$,
where $\epsilon(\mathbf{k})=s_0+s_1(k_x^2+k_y^2)$ and $\epsilon'(\mathbf{k})=s'_0+s'_1(k_x^2+k_y^2)$. $s_0,s_1,s'_0,s'_1$, and $t_1$ are parameters. $t_1$ term comes from the hopping between Os$_1$ and Os$_2$ and is essential for converting singlet pairing to triplet pairing.
% If $s_1<s'_1<0$, the structure of the bands consist with the numerical results qualitatively.
With the superconducting gap, the BdG Hamiltonian is
\begin{eqnarray}
  H(\mathbf{k})_{\mathrm{BdG}}=\left(
  \begin{array}{cc}
  h(\mathbf{k})-\mu\sigma^0 & i\Delta\sigma^y \\
  -i\Delta^*\sigma^y & -h^*(-\mathbf{k})+\mu\sigma^0
  \end{array}\right),
\end{eqnarray}
where $\Delta$ is the superconducting gap, and $\mu$ is the chemical potential. When $\mu$ is tuned to cross only one band, we can further project this $4\times4$ Hamiltonian into a $2\times2$ Hamiltonian and the $s$-wave pairing will be transformed into a $p_x+ip_y$ pairing due to the $t_1$ term in $h(\mathbf{k})$. Thus, we conclude that CTSc phases can be realized in our system by applying a gate voltage.

\begin{eqnarray}
  h_{\pm}(\mathbf{k})=\left(
  \begin{array}{cc}
  \epsilon'\pm \Delta & t_1(k_x-ik_y) \\
  t_1(k_x+ik_y) & -\epsilon'\mp \Delta
  \end{array}\right),
\end{eqnarray}

\begin{figure} %[h]
\includegraphics[width=8.6cm]{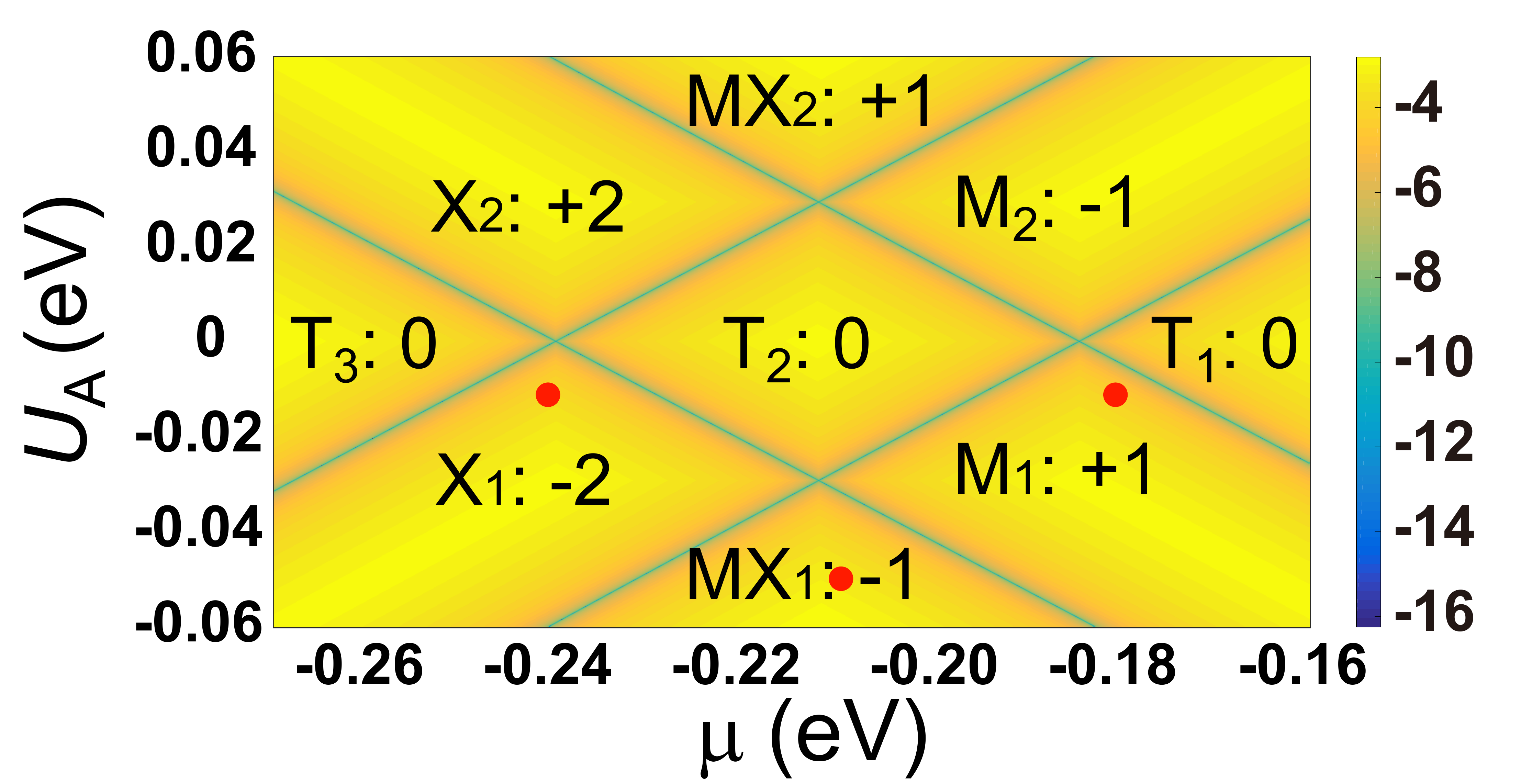}
\caption{(a) Phase diagram of the CTSc phases in the region $-0.06$ eV$<U_{\mathrm{A}}<0.06$ eV and $-0.27$ eV$<\mu<-0.16$ eV. The dark lines are phase boundaries. The number in each phase denotes the corresponding chirality. The value of $\Delta$ is fixed at $0.015$ eV.}   \label{Fig:4}
\end{figure}

The advantage of our system is that a rich phase diagram for CTSc with different Chern number $N$ can be obtained by tuning $U_{\mathrm{A}}$ and $\mu$. In Fig. \ref{Fig:4}, we identify the phase diagram in the parameter region $-0.06$ eV$<U_{\mathrm{A}}<0.06$ eV and $-0.27$ eV$<\mu<-0.16$ eV. The phase diagram is constructed by tracking the gap at X and M, and dark lines correspond to the gapless points for topological phase transitions. We have carefully checked the band gap in the whole BZ to make sure that gap closing only occurs at X and M in this parameter region. In Fig. \ref{Fig:4}, the regions labeled by $T_i(i=1,2,3)$ are for trivial phases with Chern number $N=0$, while the regions with $X_i$, $M_i$ and $MX_i$ ($i=1,2$) are for CTSc phases with Chern number $N$. We notice that the CTSc phases with the same $\mu$ but opposite $U_{\mathrm{A}}$ carry opposite Chern number $N$. Besides the CTSc $M_i: N=\pm 1$ phase discussed above, we find another two CTSc phases: $MX_i: N=\pm 1$ and $X_i: N=\pm 2$. The energy dispersions for the phase $MX_i: N=\pm 1$ ($X_i: N=\pm 2$) are shown in Fig. \ref{Fig:3}c (Fig. \ref{Fig:3}e) for the bulk and Fig. \ref{Fig:3}d (\ref{Fig:3}f) for the edge with $U_{\mathrm{A}}=-0.05$ eV and $\mu=-0.21$ eV ($U_{\mathrm{A}}=-0.01$ eV and $\mu=-0.24$ eV). Due to the multiple crossings around both X and M, chiral Majorana edge modes and other trivial edge modes coexist at the boundary in these two CTSc phases.

\section{Discussion and conclusion}
%{\it Discussion and conclusion}--
The rich phase diagram in Fig. \ref{Fig:4} provides us a platform to realize various TSCs junctions for transport experiments. For example, by tuning gate voltage between the $MX_i: N=\pm 1$ phase and the $X_i: N=\pm 2$ phase, we will be able to construct junction structures with CTSc phase with $N=1$ and the QAH phase (the $X_i: N=\pm 2$ phase is equivalent to a QAH phase). This type of junction structures will allow us to probe half-integer Hall conductivity \cite{PhysRevB.84.060510} or to realize Majorana interferometry \cite{fu2009probing,akhmerov2009electrically}.
%({\bf possible choices of the s wave superconductors by checking the lattice constant})
Finally, we emphasize that our strategy can also be generalized to other AFM materials and the key here is to break the $\hat{T}\hat{P}$ symmetry. SC proximity effect has been intensively studied in ferromagnetic materials \cite{buzdin2005proximity, bergeret2005odd} while anti-ferromagnetism is more compatible with SC. Thus,
our work suggests a new avenue to search for CTSc phases.

\section{Acknowledgements}
%{\it Acknowledgement}
X.-Y. Dong acknowledges the support from the Program of Basic Research Development of China (Grant No. 2011CB921901) and National Natural Science Foundation of China (Grant No 11374173). C.-X. Liu acknowledges the support from
Office of Naval Research (Grant No. N00014-15-1-2675).

\begin{appendix}
\section{Supplementary information for calculation details}\label{App:A}
Density-functional theory (DFT) calculations were carried out using a plane-wave basis set based on a pseudopotential framework as implemented in the Vienna {\it Ab-initio} Simulation Package (VASP)\cite{PhysRevB.47.558}. The exchange-correlation functional was the generalized gradient approximation (GGA) implemented following the Perdew--Burke--Ernzerhof prescription\cite{PhysRevLett.77.3865}. The missing correlation effect beyond GGA is taken into account through GGA$+$U calculations\cite{PhysRevB.48.16929,PhysRevB.57.1505}. Spin-orbit coupling was included for fully relativistic calculations. For the plane-wave calculation, a $600$ eV plane-wave cut-off was used. A $k$-point mesh of $8\times8\times6$ was used for the integration in the in the Brillouin zone. We adopted the lattice parameters and symmetry (I4/m) of the crystal structure according to the experimental measurement measured at $78$ K\cite{paul2013synthesis,PhysRevLett.111.167205}, for the DFT calculations. In addition of that we use experimental high temperature antiferromagnetic (AFM) configuration (called AF1), as described in the main text for all DFT self-consistent calculations. In previous experiment\cite{paul2013synthesis,PhysRevLett.111.167205}, two AFM states, referred to as AF1 and AF2, were observed. When cooling down from the room temperature, the compound exhibits a transition from the paramagnetic phase to an AFM phase (i.e. AF1) at 140 K, and a further transition to a new AFM phase (i.e. AF2) at 67 K. Along the $c$ axis, an Fe-Os double-layer (DL), in which couples in an FM way between the Fe layer and the Os layer, forms a principle stacking unit. In AF1, the DL couples to other DLs in an FM way along c. In contrast, in AF2 the Fe-Os DLs couple to each other in an AFM way along $c$. Therefore, we adopt a single DL in the thin film model, which is equivalent regarding the bulk AF1 and AF2 phases, as shown in Fig. 1(b) in the main text. To obtain the effective Hamiltonian of this double layer, we performed DFT calculations on the bulk AF1 phase and constructed the thin film model using Wannier functions (see below). Consequently, our thin film model fully considers the material structure, band structure and magnetic properties of {\it ab-initio} DFT calculations.
\setcounter{figure}{0}
\setcounter{table}{0}
\renewcommand\thefigure{S\arabic{figure}}
\renewcommand\thetable{S\arabic{table}}

\begin{figure} [h]
\includegraphics[width=8.6 cm]{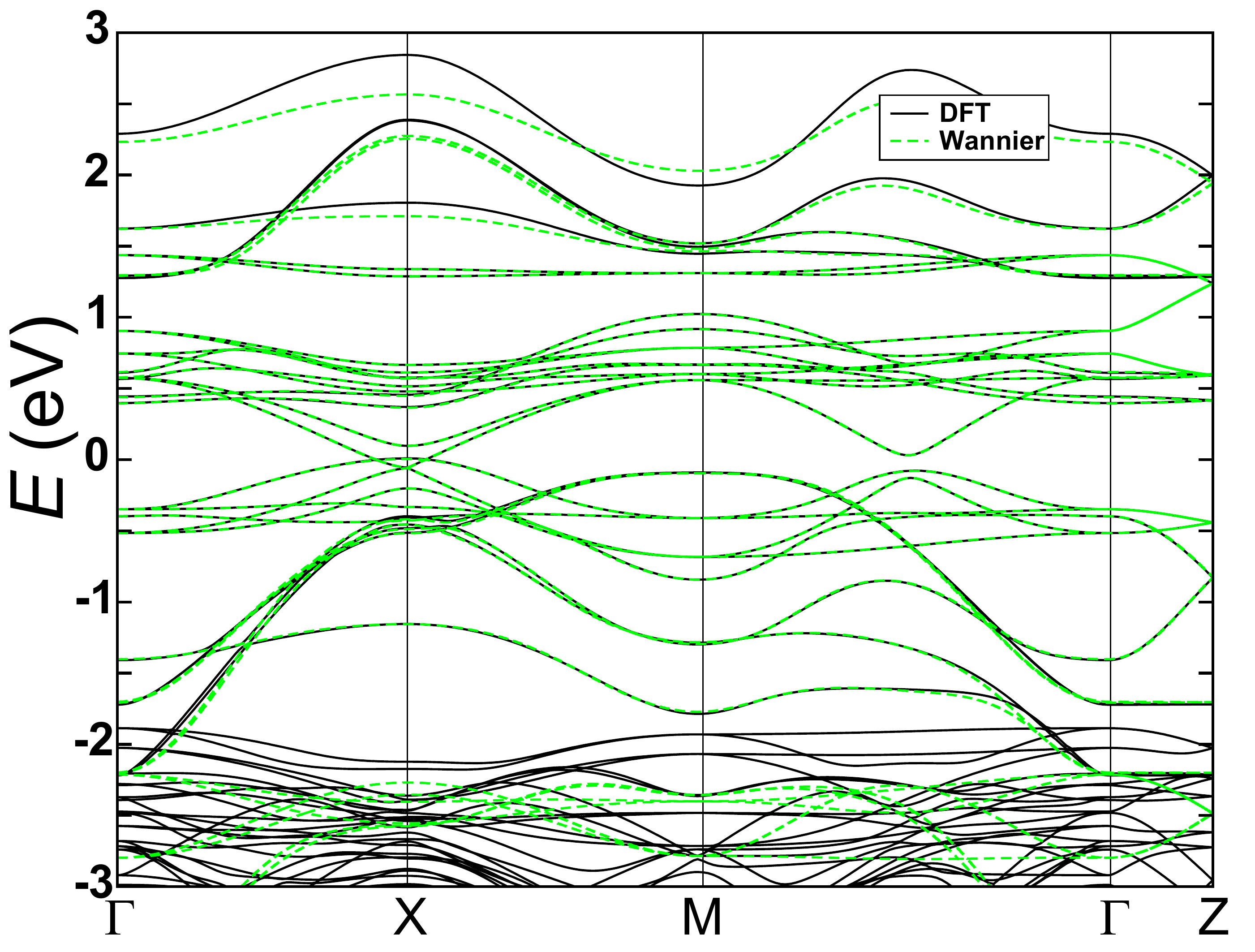}
\caption{Band structure comparison between low energy few bands keeping only Fe-$d$ and Os-$t_{2g}$ orbitals in Wannier basis (light dotted ) and full DFT bands (black continuous) calculated using GGA$+$U$_{\mathrm{eff}}$ in the antiferromagnetic configurations. U$_{\mathrm{eff}}^{\mathrm{Fe}}=4$ eV and U$_{\mathrm{eff}}^{\mathrm{Os}}=2$ eV used for evaluating the low energy Hamiltonian.}   \label{Fig:S2}
\end{figure}

In order to gain insights on the computed electronic structure, we have employed a first-principles based downfolding technique for constructing low energy Hamiltonian. This technique creates few-band, low-energy Hamiltonians from a full DFT Hamiltonians through energy-selective procedure of integrating out degrees of freedom which are not of interest. The effect of the orbitals that are integrated out has been taken into account by renormalization method in the low energy few-band Hamiltonian. If the chosen low-energy bands form an isolated set of bands, the underlying orbitals span the same Hilbert space of the corresponding Wannier functions, giving rise to Wannier functions generated in a direct manner\cite{mostofi2008wannier90}. The method provides a first-principles way for deriving a few-band, tight-binding Hamiltonian of the form $H_{\mathrm{TB}}=\sum_{ij}t^{m,n}_{i,j}(c^{\dag}_{i,m} c_{j,n}+h.c.)$ for a complex system, where the parameters $t^{m,n}_{i,j}$ are obtained numerically and define the effective hopping between the active, non-downfolded orbitals, $m$ and $n$. $c^{\dag}_{i,m}$ ($c_{j,n}$) are electron creation (annihilation) operators on site $i$ ($j$) at orbital $m$ ($n$). These downfolding calculations are based on the self-consistent potential and wave functions that obtained from the full {\it ab-initio} DFT calculations with all material specific inputs such as experimental crystal structure, chemical information, for example, charge and valence state, hybridization effect between different atoms, magnetic configuration (AF1) and exchange interactions, as well as other information such as the strength of electronic correlation and spin-orbit coupling. For the present case, the effective low-energy Hamiltonian are constructed in Wannier basis of Fe-$t_{2g}$, Fe-$e_g$ and Os-$t_{2g}$ orbitals only and integrating out all other degrees of freedom. We emphasize that the effect of oxygen $p$ orbitals and Sr orbitals is included in a renormalized manner into the few band tight binding Hamiltonian. By including the spin freedom, the low energy Hamiltonian is in the form of $32\times 32$ matrix (two Fe sites with five $d$ orbitals each site, two Os sites and three $d$ orbitals each site, and two spins for each orbital). The figure below shows that the band structure obtained by numerically solving the energy selective downfolding Hamiltonian is fully consistent with DFT band structure in the entire Brillouin zone (Fig. \ref{Fig:S2}) and such a consistence ensures that our Wannier function based Hamiltonian correctly mimicking the real dull band structure effect with full material specific information, e.g. band structure, strength of hybridization and exchange interactions, electron-electron correlation, magnetic structure and spin-orbit coupling effect. In particular, although orbitals from oxygen atoms have been integrated out through the procedure described above, their main effect to low energy physics has been included in the exchange coupling between electrons and magnetic moments, which is automatically taken into account in the diagonal term (on-site energy) of our tight-binding Hamiltonian $H_{\mathrm{TB}}$, namely $t_{ii}^{m,m}$ taking different values for opposite spins. Furthermore, we stress that this Hamiltonian does not include any empirical parameters.

\section{The effects of CDW/SDW}\label{App:B}

Above some critical gate voltage $U_{\mathrm{A}}$, the valence band top at M should be above the conduction band bottom at X (and Y). As a consequence, the electrons will move from to M to X(Y), forming an electron pocket around X(Y) and a hole pocket at M. The scattering between electron Fermi pockets around X and Y may leads to a $(\pi,\pi)$ CDW or SDW in this system. Nevertheless, the topological nature of this system (Berry curvature and Chern number) can not be immediately destroyed by the CDW or SDW since one always needs a band gap closing to induce a topological phase transition can only be induced when there is a (direct) band gap closing. In this section of the supplementary material, we will demonstrate that the main role of CDW/SDW is to split the degeneracy of energy bands at X and Y due to Brillouin zone (BZ) folding. As a consequence, the topological phase transition from Chern number $0$ to $+2$ will be split into two transitions, one from $0$ to $+1$ and the other from $+1$ to $+2$. Therefore, topological phases with non-zero Chern number remains robust for a large gate voltage.

\begin{figure} [h]
\includegraphics[width=8.6 cm]{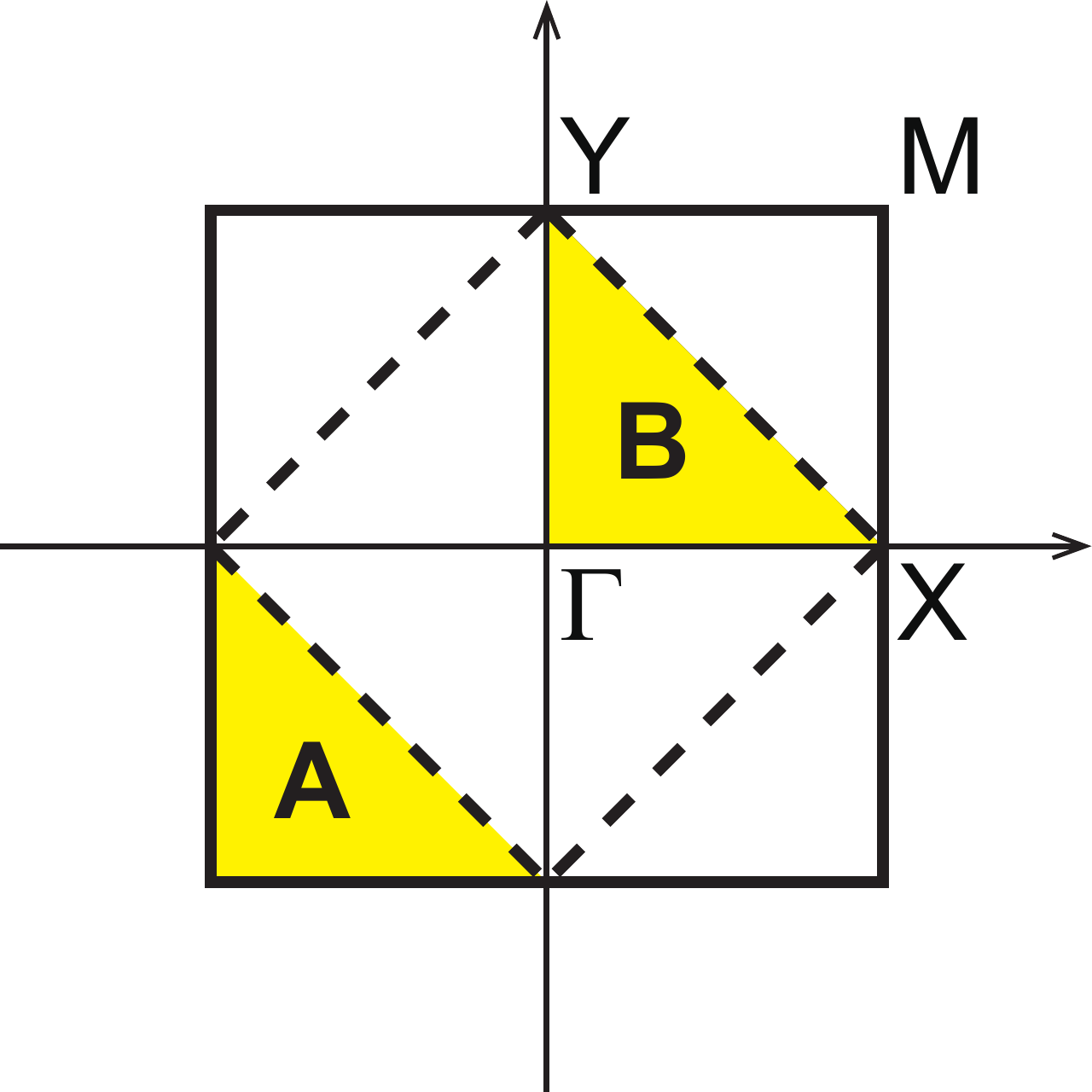}
\caption{The original BZ and the reduced BZ (enclosed by dashed lines) when there is a CDW/SDW term. The region A will move to region B during the BZ folding process.}   \label{Fig:S3}
\end{figure}

The existence of CDW or SDW will double the unit cell in real space, and the corresponding BZ will be folded. In Fig. \ref{Fig:S3}, the dashed lines enclose the reduced BZ, and the BZ folding process can be exemplified by moving the region A through $(\pi,\pi)$ to region B. In the reduced BZ, the point X and Y are connected by a reciprocal lattice vector and actually the same point.

\begin{figure} [h]
\includegraphics[width=8.6cm]{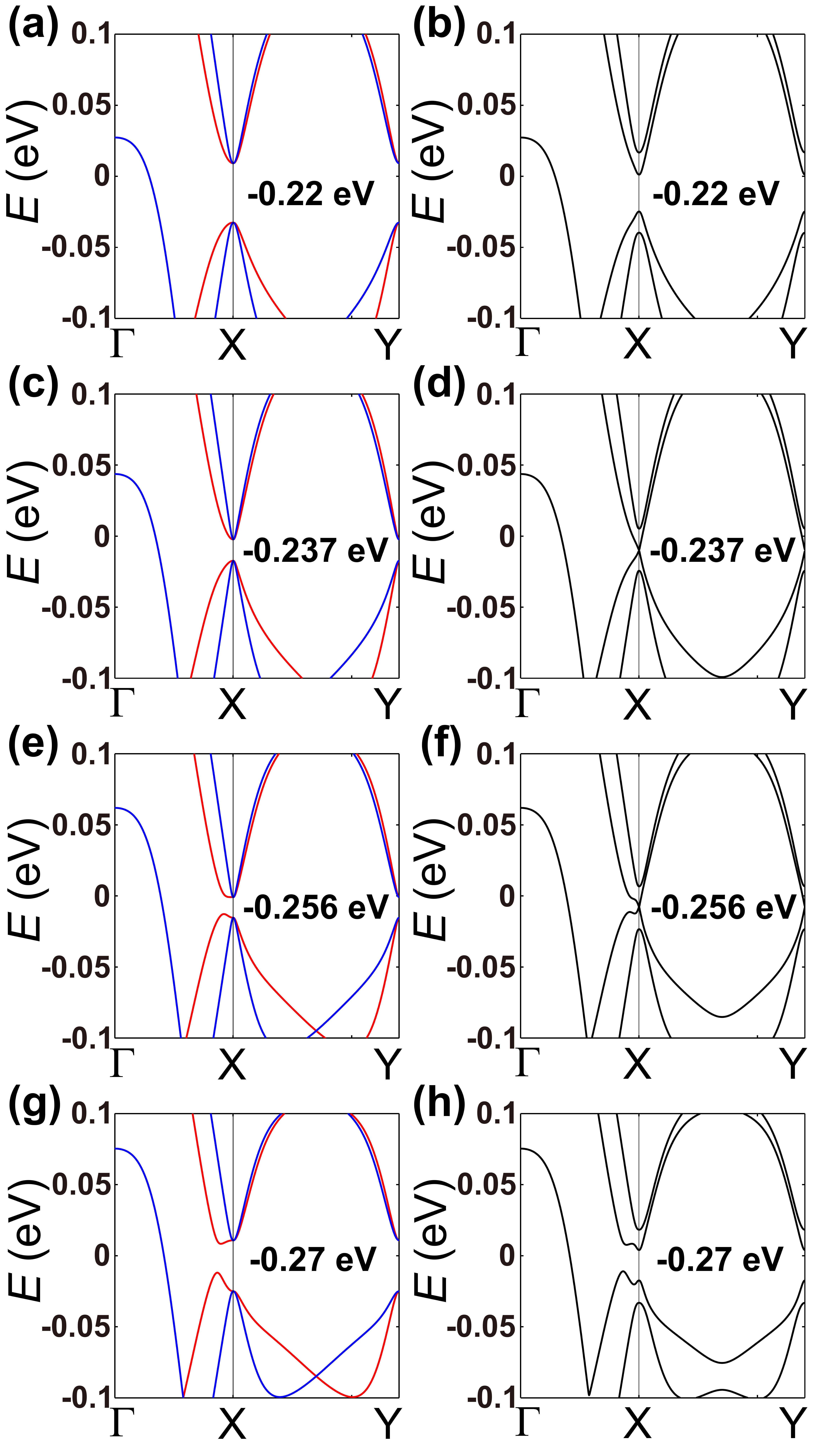}
\caption{The energy bands in the reduced BZ with different gate voltage $U_{\mathrm{A}}$ (the values in each figure). The blue and red lines in (a), (c), (e), and (g) are the bands without CDW/SDW term. The black lines in (b), (d), (f), and (h) are the bands with CDW/SDW term.}   \label{Fig:S4}
\end{figure}

The band structures with different gate voltage $U_{\mathrm{A}}$ are shown in the reduced BZ on $\Gamma$-X-Y in Fig. \ref{Fig:S4}. The blue and red lines in the left penal (Fig. \ref{Fig:S4}(a, c, e, g)) are the energy bands without CDW/SDW term, while the black lines in the right penal (Fig. \ref{Fig:S4}(b, d, f, h)) are the energy bands with a small CDW/SDW term. We know that without CDW/SDW term the energy at X and Y point are the same, and in the reduced BZ there will be a double degeneracy at X and Y. As is shown in the main text, the band gap at X (Y) closes when $U_{\mathrm{A}}=-0.246$ eV, and the Chern number of the system changes from $0$ to $+2$. When the CDW/SDW term exists, this double degeneracy will be broken and the bonding and anti-bonding states between the electron pockets at X and Y will be formed. In this case, if we tune the gate voltage, two band inversions at X point will happen successively when $U_{\mathrm{A}1}=-0.237$ eV (see Fig. \ref{Fig:S4}(d)) and $U_{\mathrm{A}2}=-0.256$ eV (see Fig. \ref{Fig:S4}(f)). For the gate voltage between $U_{\mathrm{A}1}$ and $U_{\mathrm{A}2}$, the Chern number of the system is $+1$, since only one band inversion happens. While for $U_{\mathrm{A}}<U_{\mathrm{A}2}$  (see Fig. \ref{Fig:S4}(h)), two band inversions take place and the Chern number will be +2. Therefore, we have proved that the possible CDW/SDW term will not destroy the scenario of the Dirac physics, and more anomalous Hall phase such as the phase with Chern number $\pm 1$ could emerge.

\section{anomalous Hall phase in a four-layers film}\label{App:C}
\begin{figure} [h]
\includegraphics[width=8.6cm]{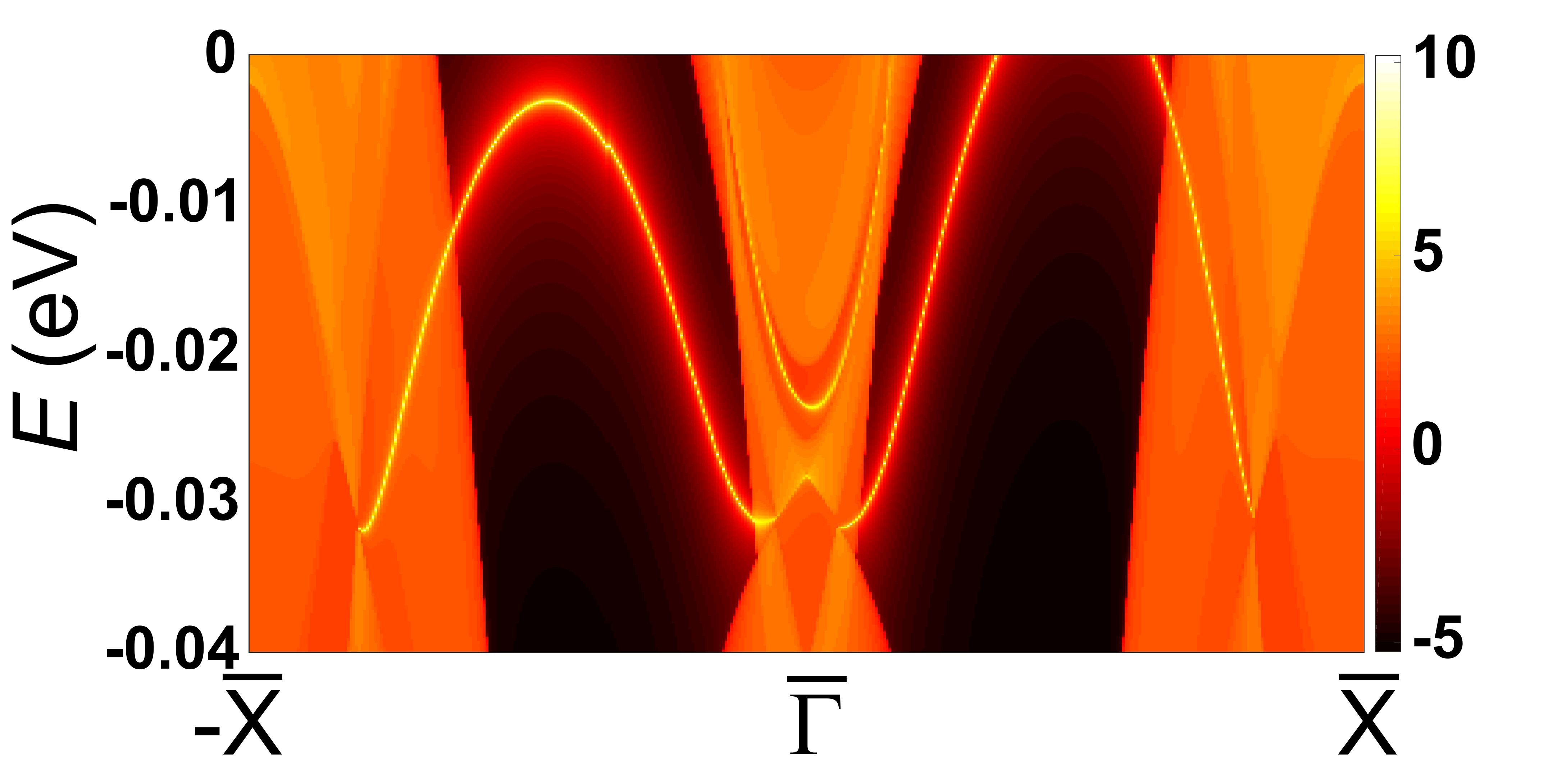}
\caption{The density of states on one edge of a ribbon configuration along $x$ direction of a four-layers film. The effective electric potential on the top layer is $-0.23$ eV and on the bottom layer is $0.23$ eV.}   \label{Fig:S1}
\end{figure}

The edge dispersion of a four-layers film in the ribbon configuration is shown in Fig. \ref{Fig:S1}. We add opposite effective electric potential on the top ($-0.23$ eV) and bottom layer ($0.23$ eV). In the four-layers film the edge modes appears with a smaller asymmetric potential compared to the bilayer case ($U_{\mathrm{A}}<-0.246$ eV) in the main text.

\section{Effective model around M point}\label{App:D}
In this section, we will construct the effective model around M point for our system based on symmetry argument.
Without external gate voltage and superconducting pairing term, the highest valence band around M point in the BZ is doubly degenerate. We denote the wave function as $|\psi_1\rangle_{\mathrm{M}}$ and $|\psi_2\rangle_{\mathrm{M}}$. Since the double degeneracy comes from the existence of the symmetry $\hat{T}\hat{P}$, these two states must be related to each other as $|\psi_1\rangle_{\mathrm{M}}=\hat{T}\hat{P}|\psi_2\rangle_{\mathrm{M}}$. With the help of the numerical calculations, we find
\begin{eqnarray}
  |\psi_1\rangle&=&f_0^*|\mathrm{Os}_1,d_{xy}, \uparrow\rangle +f_1^*|\mathrm{Fe}_1,d_{z^2},\uparrow\rangle\nonumber\\
  && +f_2^*(|\mathrm{Os}_1,d_{xz},\downarrow\rangle -i|\mathrm{Os}_1,d_{yz},\downarrow\rangle)\nonumber\\
  && +f_3^*(|\mathrm{Os}_2,d_{xz},\downarrow\rangle +i|\mathrm{Os}_2,d_{yz},\downarrow\rangle)
\end{eqnarray}
and
\begin{eqnarray}
  |\psi_2\rangle&=&-f_0|\mathrm{Os}_2,d_{xy}, \downarrow\rangle -f_1|\mathrm{Fe}_2,d_{z^2},\downarrow\rangle\nonumber\\
  && +f_2(|\mathrm{Os}_2,d_{xz},\uparrow\rangle +i|\mathrm{Os}_2,d_{yz},\uparrow\rangle)\nonumber\\
  && +f_3(|\mathrm{Os}_1,d_{xz},\uparrow\rangle -i|\mathrm{Os}_1,d_{yz},\uparrow\rangle)
\end{eqnarray}
which satisfy the relation $\hat{T}\hat{P}|\psi_1\rangle=|\psi_2\rangle$, and
\begin{eqnarray}
  \hat{T}\hat{P}|\psi_2\rangle=(\hat{T}\hat{P})^2|\psi_1\rangle
  =-|\psi_1\rangle.
\end{eqnarray}

As a result, on the basis $\Psi=(|\psi_1\rangle,|\psi_2\rangle)^T$, the matrix representation of $\hat{T}\hat{P}$ is written as
\begin{eqnarray}
  D(\hat{T}\hat{P})=\left(
  \begin{array}{cc}
    0 & -1\\
    1 & 0\\
  \end{array}\right)
\end{eqnarray}

Additional three symmetries, $\hat{T}\hat{m}_x$, $\hat{T}\hat{m}_y$, and $\hat{c}_4(z)$ (four-fold rotation symmetry around $z$ axis), should be taken into account. Since under the symmetry operations the behaviors of $|\psi_{1(2)}\rangle$ is the same with $|\mathrm{Os}_{1(2)},d_{xy},\uparrow(\downarrow)\rangle$, we investigate the transformations of the later ones under the symmetry operations.

We choose the reflection plane of $\hat{m}_x$ and $\hat{m}_y$ as the plane consisting of $\mathrm{Os}_1$ and $\mathrm{Fe}_2$ and find
\begin{eqnarray}
  &&\hat{T}\hat{m_x}|\mathrm{Os}_1,d_{xy},\uparrow\rangle (\mathbf{k}) \nonumber\\
  &&=\hat{T}\hat{m_x}\frac{1}{\sqrt{N}}\sum_{\mathbf{R}} e^{i\mathbf{k}\cdot\mathbf{R}}\phi_{dxy}(\mathbf{r} -\mathbf{R}-\mathbf{r}_{\mathrm{Os}_1})|\uparrow\rangle \nonumber\\
  &&=i|\mathrm{Os}_1,d_{xy},\uparrow\rangle (-\hat{m}_x\mathbf{k})
\end{eqnarray}
and
\begin{eqnarray}
  &&\hat{T}\hat{m_x}|\mathrm{Os}_2,d_{xy},\downarrow\rangle (\mathbf{k})=-ie^{ik_x}|\mathrm{Os}_2,d_{xy},\downarrow\rangle (-\hat{m}_x\mathbf{k}).
\end{eqnarray}

With $\hat{T}\hat{m}_x|\psi_i(\mathbf{k})\rangle =\sum_jD(\hat{T}\hat{m}_x)_{ji}|\psi_j(-\hat{m}_x\mathbf{k}) \rangle$, the
matrix form of $\hat{T}\hat{m}_x$ is given by
\begin{eqnarray}
  D(\hat{T}\hat{m}_x)=\left(
  \begin{array}{cc}
    i & 0\\
    0 & -ie^{ik_x}\\
  \end{array}\right).
\end{eqnarray}
At $M=(\pi,\pi)$, we have $D(\hat{T}\hat{m}_x)=i\sigma^0$ and $\hat{T}\hat{m}_x=i\sigma^0K$.

For $\hat{T}\hat{m}_y$,
\begin{eqnarray}
  &&\hat{T}\hat{m_y}|\mathrm{Os}_1,d_{xy},\uparrow\rangle (\mathbf{k})=|\mathrm{Os}_1,d_{xy},\uparrow\rangle (-\hat{m}_y\mathbf{k})
\end{eqnarray}
and
\begin{eqnarray}
  &&\hat{T}\hat{m_y}|\mathrm{Os}_2,d_{xy},\downarrow\rangle (\mathbf{k})=e^{ik_y}|\mathrm{Os}_2,d_{xy},\downarrow\rangle (-\hat{m}_y\mathbf{k}).
\end{eqnarray}

With $\hat{T}\hat{m}_y|\psi_i(\mathbf{k})\rangle =\sum_jD(\hat{T}\hat{m}_y)_{ji}|\psi_j(-\hat{m}_y\mathbf{k}) \rangle$, we have
\begin{eqnarray}
  D(\hat{T}\hat{m}_y)=\left(
  \begin{array}{cc}
    1 & 0\\
    0 & e^{ik_y}\\
  \end{array}\right),
\end{eqnarray}
leading to $D(\hat{T}\hat{m}_y)=\sigma^z$ and $\hat{T}\hat{m}_y=\sigma^zK$ at $M=(\pi,\pi)$.

For $\hat{c}_4(z)$, we get
\begin{eqnarray}
  &&\hat{c}_4|\mathrm{Os}_1,d_{xy},\uparrow\rangle (\mathbf{k}) =(-1)\frac{1+i}{\sqrt{2}} |\mathrm{Os}_1,d_{xy},\uparrow\rangle (\hat{c}_4\mathbf{k})
\end{eqnarray}
and
\begin{eqnarray}
  &&\hat{c}_4 |\mathrm{Os}_2,d_{xy},\downarrow\rangle (\mathbf{k})=-e^{-ik_x}\frac{1-i}{\sqrt{2}} |\mathrm{Os}_2,d_{xy},\downarrow\rangle (\hat{c}_4\mathbf{k}).
\end{eqnarray}
With $\hat{c}_4|\psi_i(\mathbf{k})\rangle =\sum_jD(\hat{c}_4)_{ji}|\psi_j(-\hat{m}_y\mathbf{k}) \rangle$, we have
\begin{eqnarray}
  D(\hat{c}_4)=\left(
  \begin{array}{cc}
    -\frac{1+i}{\sqrt{2}} & 0\\
    0 & \frac{1-i}{\sqrt{2}}\\
  \end{array}\right) =\frac{-1}{\sqrt{2}}(i\sigma^0+\sigma^z)
\end{eqnarray}

In summary, we have
\begin{eqnarray}
  &&\hat{T}\hat{P}=-i\sigma^yK\\
  &&\hat{T}\hat{m}_x=i\sigma^0K\\
  &&\hat{T}\hat{m}_y=\sigma^zK\\
  &&\hat{c}_4=\frac{-1}{\sqrt{2}}(i\sigma^0+\sigma^z).
\end{eqnarray}

The transformation properties of the Pauli matrix and $\mathbf{k}$ are summarized in Tab. \ref{tab:1}.

\begin{table} [h]
\centering
\begin{tabular}{c|c|cccc||c}
    \hline\hline
      &  & $\hat{T}\hat{P}$ & $\hat{T}\hat{m}_x$ & $\hat{T}\hat{m}_y$ & $\hat{c}_4(z)$ & \\
    \hline
    $\sigma^0$ & $c,k_x^2+k_y^2$ & + & + & + & +& $c,k_x^2+k_y^2$\\
    $\sigma^x$ & $N$ & - & + & - &  & $N$\\
    $\sigma^y$ & $N$ & - & - & + &  & $N$\\
    $\sigma^x+i\sigma^y$ & $N$ &   & + & - & $-i$ & $k_x+ik_y$\\
    $\sigma^x-i\sigma^y$ & $N$ &   & + & - & $+i$ & $k_x-ik_y$\\
    $\sigma^z$ & $N$ & - & + & + & + &$c,k_x^2+k_y^2$\\
    \hline
  \end{tabular}
\caption{Symmetry analysis. Transformation properties of the $\sigma_i$ matrices and the polynomials of $\mathbf{k}$ up to the second order under the symmetry operations, where $\mathbf{k}$ is measured relative to M point in the BZ. $N$ in the table means there is no polynomials of $\mathbf{k}$ up to the second order satisfies the corresponding transformation rule. The $\mathbf{k}$ in the second column satisfy all the symmetries. The $\mathbf{k}$ in the last column satisfy only $\hat{T}\hat{m}_x$, $\hat{T}\hat{m}_y$, and $\hat{c}_4(z)$. $\hat{c}_4(z)$ changes $(k_x,k_y)$ to $(k_y,-k_x)$. }\label{tab:1}
%  \end{minipage}
\end{table}

When there is no gate voltage, the $\hat{T}\hat{P}$ symmetry exists and the only possible term in the two band effective Hamiltonian is $(c_0+c_1(k_x^2+k_y^2))\sigma^0$, giving rise to a doubly degenerate parabolic band dispersion around M point in the BZ. For a non-zero gate voltage, the $\hat{T}\hat{P}$ symmetry will be broken and the possible form of the two band Hamiltonian is written as
\begin{eqnarray}
  h(\mathbf{k})&=&\epsilon(\mathbf{k})\sigma^0 +\epsilon'(\mathbf{k})\sigma^z+t_1(k_x+ik_y)(\sigma^x-i\sigma^y)/2\nonumber\\
  &&+t_1(k_x-ik_y)(\sigma^x+i\sigma^y)\nonumber\\
  &=&\epsilon(\mathbf{k})\sigma^0 +\epsilon'(\mathbf{k})\sigma^z+t_1(k_x\sigma^x+k_y\sigma^y)
\end{eqnarray}
where $\epsilon(\mathbf{k})=s_0+s_1(k_x^2+k_y^2)$ and $\epsilon'(\mathbf{k})=s'_0+s'_1(k_x^2+k_y^2)$.

Next we will illustrate how the $t_1$ term in the Hamiltonian $h(\mathbf{k})$ emerges from the view point of $\mathbf{k}\cdot\mathbf{p}$ method. Since the $t_1$ term is linear, it should originate from the non-zero term of $\langle\psi_1|{\bf k}\cdot\mathbf{p}|\psi_2\rangle$. According to
the form of $|\psi_1\rangle$ and $|\psi_2\rangle$,
the off-diagonal elements of the effective model $h(\mathbf{k})$ could come from the hopping between the Os$_1$ and Os$_2$, such as the term $Q_1=\langle\mathrm{Os}_1,d_{xy}, \uparrow|(p_x-ip_y)(|\mathrm{Os}_2,d_{xz},\uparrow\rangle +i|\mathrm{Os}_2,d_{yz},\uparrow\rangle)$.
 We need to check the symmetry constraint on these matrix elements. Let's take the $\hat{c}_4(z)$ symmetry as an example.
 We may take the eigen form of the wave functions and operators. For example, the wave functionf $(|\mathrm{Os}_2,d_{xz},\uparrow\rangle +i|\mathrm{Os}_2,d_{yz},\uparrow\rangle)$ transforms with the eigen-value $(1-i)/\sqrt{2}$ under $\hat{c}_4(z)$ at M point, the wave function $|\mathrm{Os}_1,d_{xy}, \uparrow\rangle$ under $\hat{c}_4(z)$ transforms with the eigen-value $-(i+1)/\sqrt{2}$ and the
 operator $(p_y-ip_x)$ transforms with eigen-value $-i$. The combination of these two wave functions and one operator can give rise to identity
 representation and thus the matrix element $Q_1$ is allowed by symmetry (selection rule).  Similar procedure can be applied to other matrix elements, leading to the following non-zero matrix elements
\begin{eqnarray}
  &&P_1=\langle\mathrm{Os}_1,d_{xy}, \uparrow|p_-(|\mathrm{Os}_2,d_{xz},\uparrow\rangle +i|\mathrm{Os}_2,d_{yz},\uparrow\rangle)\nonumber\\
  &&P_2=\langle\mathrm{Fe}_1,d_{z^2}, \uparrow|p_-(|\mathrm{Os}_2,d_{xz},\uparrow\rangle +i|\mathrm{Os}_2,d_{yz},\uparrow\rangle)\nonumber\\
  &&P_3=\langle\mathrm{Fe}_1,d_{z^2}, \uparrow|p_-(|\mathrm{Os}_1,d_{xz},\uparrow\rangle -i|\mathrm{Os}_1,d_{yz},\uparrow\rangle)\nonumber\\
  &&Q_1=(\langle\mathrm{Os}_1,d_{xz},\downarrow| +i\langle\mathrm{Os}_1,d_{yz},\downarrow|)p_- |\mathrm{Os}_2,d_{xy},\downarrow\rangle\nonumber\\
  &&Q_2=(\langle\mathrm{Os}_1,d_{xz},\downarrow| +i\langle\mathrm{Os}_1,d_{yz},\downarrow|)p_- |\mathrm{Fe}_2,d_{z^2},\downarrow\rangle\nonumber\\
  &&Q_3=(\langle\mathrm{Os}_2,d_{xz},\downarrow| -i\langle\mathrm{Os}_2,d_{yz},\downarrow|)p_- |\mathrm{Fe}_2,d_{z^2},\downarrow\rangle
\end{eqnarray}
where $p_-=p_x-ip_y$. As a result, $t_1$ can be expressed as
\begin{eqnarray}
  t_1&=&f_0f_2(P_1-Q_1)+f_1f_2(P_2-Q_2)+f_1f_3(P_3-Q_3)\nonumber\\
\end{eqnarray}

\end{appendix}

%\bibliography{FeOs}
%

\end{document}